\documentclass[aps,pre,twocolumn,letter]{revtex4-1}
\usepackage{graphicx}
\usepackage{float}
\usepackage{amsmath}
\usepackage{amsfonts}
\usepackage{color}
\usepackage{xr}
\usepackage{ulem}

\newcommand{\jamesemph}[1]
{\textcolor{cyan}{#1}}

\newcommand{\cn}[1]
{\jamesemph{[Citation Needed!]}}

\begin{document}
%\title{Generating Hyperuniform Jammed Configurations of Athermal Soft Spheres}
\title{Hyperuniform Jammed Sphere Packings Have Anomalous Material Properties}
\date{\today}
\author{Jack R. Dale, James D. Sartor, R. Cameron Dennis, and Eric I. Corwin}
\affiliation{Department of Physics and Materials Science Institute, University of Oregon, Eugene, Oregon 97403, USA.}

\begin{abstract}
A spatial distribution is hyperuniform if it has local density fluctuations that vanish in the limit of long length scales.  Hyperuniformity is a well known property of both crystals and quasicrystals. Of recent interest, however, is disordered hyperuniformity: the presence of hyperuniform scaling without long-range configurational order. Jammed granular packings have been proposed as an example of disordered hyperuniformity, but recent numerical investigation has revealed that many jammed systems instead exhibit a complex set of distinct behaviors at long, emergent length scales. We use the Voronoi tesselation as a tool to define a set of rescaling transformations that can impose hyperuniformity on an arbitrary weighted point process, and show that these transformations can be used in simulations to iteratively generate hyperuniform, mechanically stable packings of athermal soft spheres. These hyperuniform jammed packings display atypical mechanical properties, particularly in the low-frequency phononic excitations, which exhibit an isolated band of highly collective modes and a band-gap around zero frequency.
\end{abstract}

\maketitle

%Statement of paper: We have an algorithm to produce HU overjammed packings.  These packings show unexpected mechanical properties and also suggest that critical jamming is incompatible with HU.

%Figures
%\begin{itemize}
% \item Voronoi Rescaling Figure (Illustration and density fluctuation vs size)
% \item Structure factor scaling
% \item Mode illustration figure
% \item Mechanical Properties (fill in remainder of SCN figure, use slightly bigger data points)
%\end{itemize}

\section{Introduction}
A statistically homogeneous distribution of points throughout some volume is uniform in the sense that the probability density function is constant throughout the volume.  However, any finite sampling of this function will necessarily contain density fluctuations at all length scales.  By contrast, a hyperuniform random distribution is one in which density fluctuations are completely suppressed at long length scales \cite{torquato_local_2003, torquato_hyperuniformity_2016}.  While this phenomena is most closely associated with crystalline and quasi-crystalline patterns~\cite{zachary_hyperuniformity_2009}, hyperuniformity also appears in disordered systems such as the prime numbers~\cite{Torquato_hidden_2019, casini_short_2015} as well as in materials which are both disordered and statistically isotropic~\cite{chen_binary_2018, chen_designing_2018, tsurusawa_direct_2019, lomba_disordered_2017, weijs_emergent_2015, le_thien_enhanced_2017, tjhung_hyperuniform_2015, wilken_hyperuniform_2020, zachary_hyperuniformity_2009}. Hyperuniformity can even be realized in amorphous packings of circles and spheres~\cite{chremos_hidden_2018, wu_search_2015, zachary_hyperuniform_2011, kim_methodology_2019, yanagishima_towards_2020}. Further, hyperuniform materials could be used for creating efficient sensory networks without the drawbacks of aliasing. These hyperuniform sensory networks have even been shown to arise in nature, in particular in the distribution of photocells within the eyes of chickens \cite{jiao_avian_2014}.

It is natural to suspect that all jammed sphere packings are hyperuniform as the relaxation process by which such packings are formed will necessarily involve smoothing out local density fluctuations relative to an uncorrelated initial configuration~\cite{torquato_local_2003}.  However, recent numerical evidence demonstrates that, for typical protocols used to create jammed packings \textit{in silico}, an emergent length scale appears at which this suppression abates~\cite{wu_search_2015, ikeda_large-scale_2017}.  Further exploration of the link between hyperuniformity and jamming could give rise to deeper insights into the nature of the glass transition, and more generally of geometric order within amorphous materials. In this light, any protocol which can generate concrete examples of hyperuniform materials has significant theoretical and numerical applications.

\section{Creating Hyperuniform packings}

In order to create hyperuniform jammed packings we combine a protocol for creating mechanically stable sphere packings with a straightforward protocol for transforming any weighted point pattern into a hyperuniform point pattern. This recipe was developed in parallel with reference~\cite{yanagishima_towards_2020}.

\begin{figure}[b]
\includegraphics[width=1\linewidth]{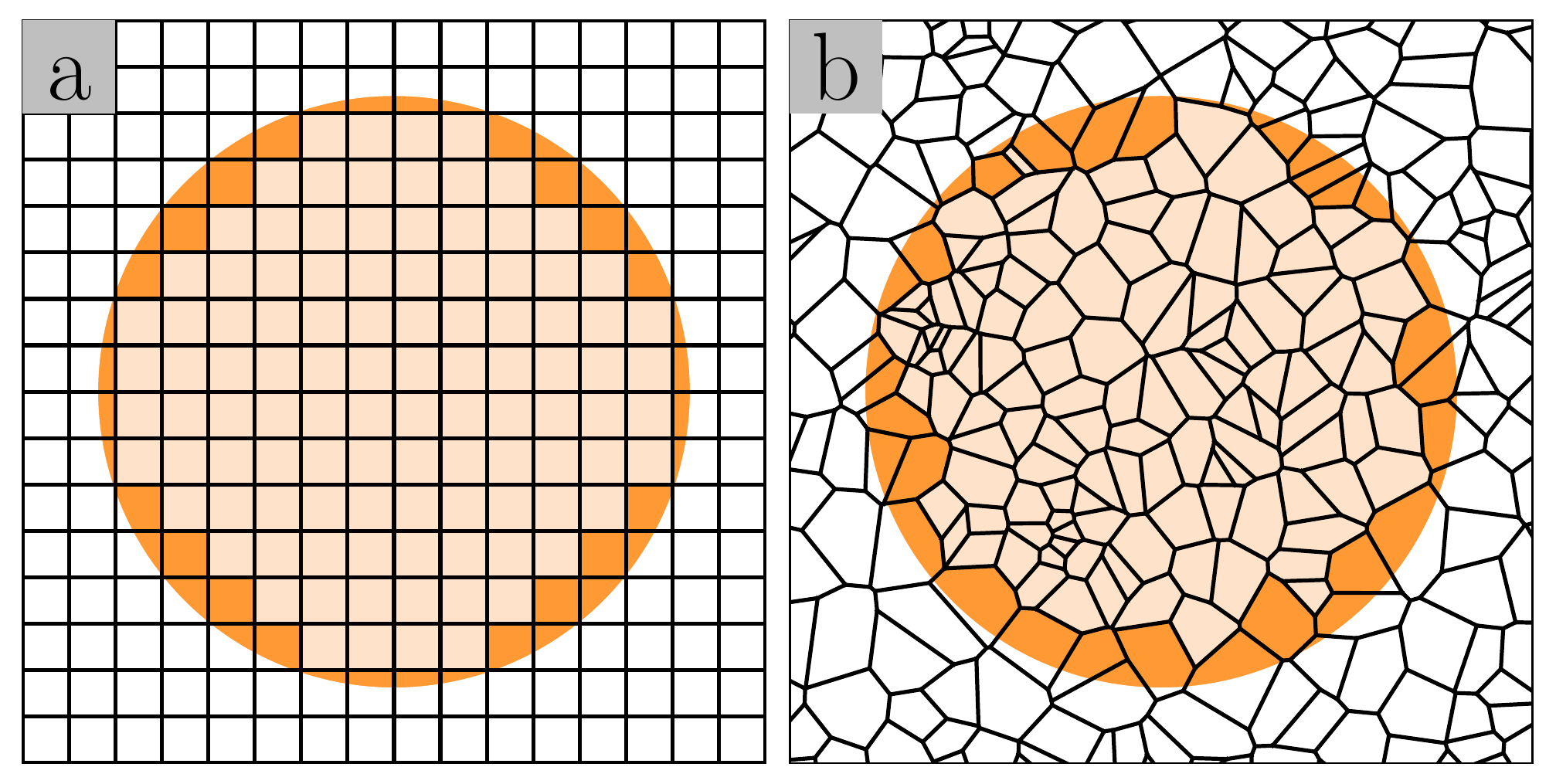}
\caption{The cells fully (light orange) and partially within (orange) a circular region. a) in a periodic tesselation, which is hyperuniform, the variance within a sampling window is introduced only from the cells on the window's boundary. b) in the tesselation of a Poisson point pattern, variance arises from variation in both the bulk and the surface of the window, leading to uniform statistics. If, however, the weight of each cell is made to be proportional to its area, then there will be no density fluctuations in the interior and as such it will follow hyperuniform statistics.}
\label{fig:HUIllustration}
\end{figure}

\begin{figure}[h]
\includegraphics[width=1\linewidth]{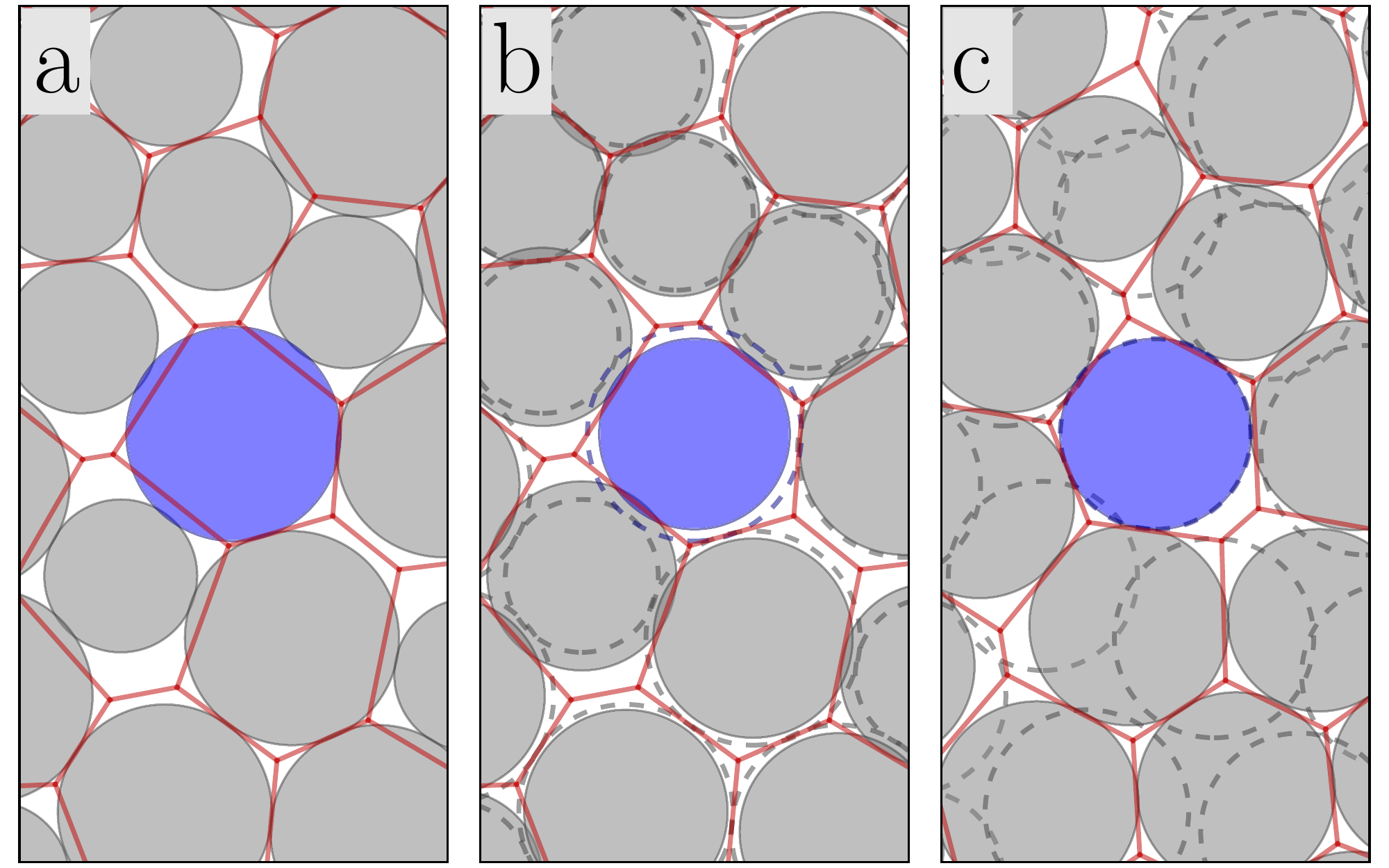}
\caption{llustration of the Voronoi rescaling. a) a mechanically stable backing is generated and its Voronoi tessellation is computed, shown in red lines. Middle: the radii are rescaled (from dashed circles to filled solid circles) so that each particle occupies a fixed fraction of its Voronoi cell, making the system hyperuniform but disrupting the mechanical equilibrium. b) The packing is relaxed (from previous positions in dashed circles to filled solid circles) back to mechanical stability.}
\label{fig:voronoiIllustration}
\end{figure}

\subsection{Mechanically Stable Sphere Packings}
Packings are simulated as sets of $N$ spherical particles each in a cubic periodic bounding box of unit volume. Particles have radii $r_i$ and positions $\vec x_i$, and interact through a harmonic pairwise interaction:
\begin{equation}
U=\sum_{i>j}\frac 12\left(1-\frac{|\vec x_j-\vec x_i|}{r_i+r_j}\right)^2\Theta\left(1-\frac{|\vec x_j-\vec x_i|}{r_i+r_j}\right)
\end{equation}
where $\Theta$ is the Heaviside step function.

Packings are created from random starting positions at a fixed packing fraction $\varphi$ and then minimized using the FIRE algorithm until the packing reaches mechanical stability~\cite{bitzek_structural_2006}. These relaxation methods are chosen to approximate a rapid quench from infinite temperature. 

\subsection{From Weighted Point Pattern to Hyperuniformity}
Given a point pattern in $d$ dimensions we may describe density fluctuations by considering the number of particles which fall within an ensemble of similarly shaped window functions.  We define $\sigma^2_l (r)$ to be the variance of the total number to fall within the subset of windows sharing a length scale $r$.  

As illustrated in figure \ref{fig:HUIllustration}, a Poisson point pattern, or indeed any sufficiently uncorrelated distribution, will have a variance which grows linearly with the mean number of points and so will have $\sigma^2_l (r) \propto r^d$.  Conversely, in any periodic or quasi-crystalline distribution the only contribution to variance within a sampling window comes from the unit cells that are bisected by the window's boundary, leading to a variance which grows as the surface area $\sigma^2_l (r) \propto r^{d-1}$. Such a point pattern is said to be hyperuniform.  This definition may be generalized to weighted point patterns by considering the variance of the total weight within the window function rather than the number of points.   For purposes of analyzing density fluctuations in granular packings, we idealize each particle as a delta distribution at the particle's center with weight equal to the volume of the particle. This choice has been shown to not affect the long range correlations \cite{wu_search_2015}. Thus we compute the density distribution $\rho$ as
\begin{equation}
\rho(\vec x)=\sum_i \frac{\pi^{d/2}}{\Gamma\left(\frac{d}{2}+1\right)} r_i^d \delta(\vec x-\vec x_i).
\end{equation}

\begin{figure}[h]
\includegraphics[width=1\linewidth]{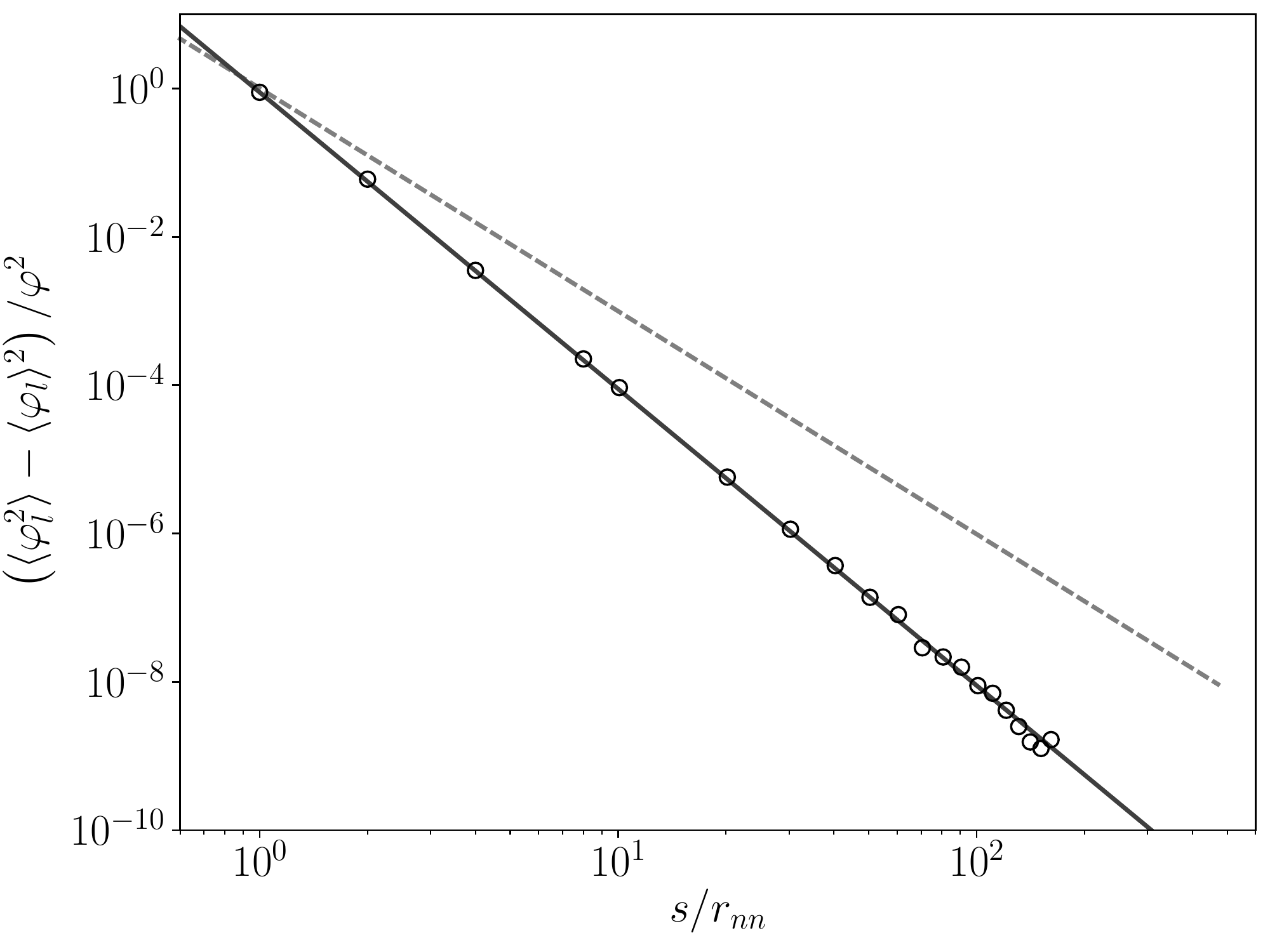}
\caption{Fluctuations in local density, $\varphi_l$, in a cubic region plotted against the side length $s$ of the cubic region scaled by the typical distance between nearest neighbors, $r_{nn}$. Black circles show the measured density fluctuations for a rescaled Poisson point distribution of $2^{25} = 33,554,432$ particles in 3 dimensions. The hyperuniform power law with exponent $-(d+1)$ is shown as a solid black line. The grey dashed line shows the power law distribution for a uniform distribution, $-d$.}
\label{fig:realSpaceScaling}
\end{figure}

We propose a method to create hyperuniform packings from conventional packings by iteratively resizing particles to achieve uniform local density in the system, as illustrated in figure \ref{fig:voronoiIllustration}.  We first turn the point pattern associated with the packing into a hyperuniform point pattern with the following algorithm:
%
%A periodic point distribution will give rise to a tesselation with a small number of repeated cells and so the only contribution to variance within a sampling window comes from the cells that are intersected by the window's boundary.  This gives rise to the surface area scaling that creates hyperuniform statistics \jamesemph{[maybe ref figure here too]}. In like fashion, any point pattern which supports a tessellation in which each cell has a fixed local density will also be hyperuniform.  Thus, given any weighted point pattern and a target density $\varphi$ we can create a new, hyperuniform, weighted point pattern at that density by the following algorithm:
\begin{enumerate}
 \item Compute the Voronoi tesselation~\cite{aurenhammer_voronoi_2013} associated with the point pattern.
 \item Compute the volume of each cell in the tesselation.
 \item Assign to each point a \textit{new} weight proportional to the volume of the local cell.
\end{enumerate}
In figure \ref{fig:realSpaceScaling}, we demonstrate the validity of this method for creating hyperuniform distributions. We do so by applying this Voronoi rescaling to a large number of Poisson points in a periodic box. We note that although we use the Voronoi tesselation, any tesselation with a well-behaved diameter distribution should work equally well. 

In the context of sphere packings this process will change each particle's volume and thus the polydispersity of the spheres. Even with a mechanically stable starting configuration, this Voronoi rescaling will generally result in a mechanically unstable system. To generate a packing that is \textit{both} hyperuniform and mechanically stable, we must find a fixed point of both rescaling and energy minimization.

To locate such a fixed point, we begin with mechanically stable monodisperse packings in $d$ = 3. The packings are then alternately Voronoi rescaled and partially minimized. %\jamesemph{[cut this unnecessary parenthetical? also, is our system thermal? i am confused (a full quench is not necessary here, since a configuration is a fixed point of an infinite-time quench if and only if it is a fixed point of a finite-time quench)]}. 
We repeat this process until the unbalanced body forces remain below our minimization threshold even after rescaling.  
We find that such a procedure will almost always converge to a fixed point that is a mechanically stable, hyperuniform packing. 
%For systems approaching the jamming point, the time to convergence diverges, and the fraction of convergent systems decreases. 

\section{Analysis}

\subsection{Convergence}

The dynamics of our iterative algorithm are not monotonic in energy, and explore a particularly high-dimensional configuaration space of positions and radii in which we expect hyperuniform fixed points to be relatively rare. It is thus surprising that this algorithm would ever converge, much less as robustly as it does. 

\begin{figure}[h]
\includegraphics[width=1\linewidth]{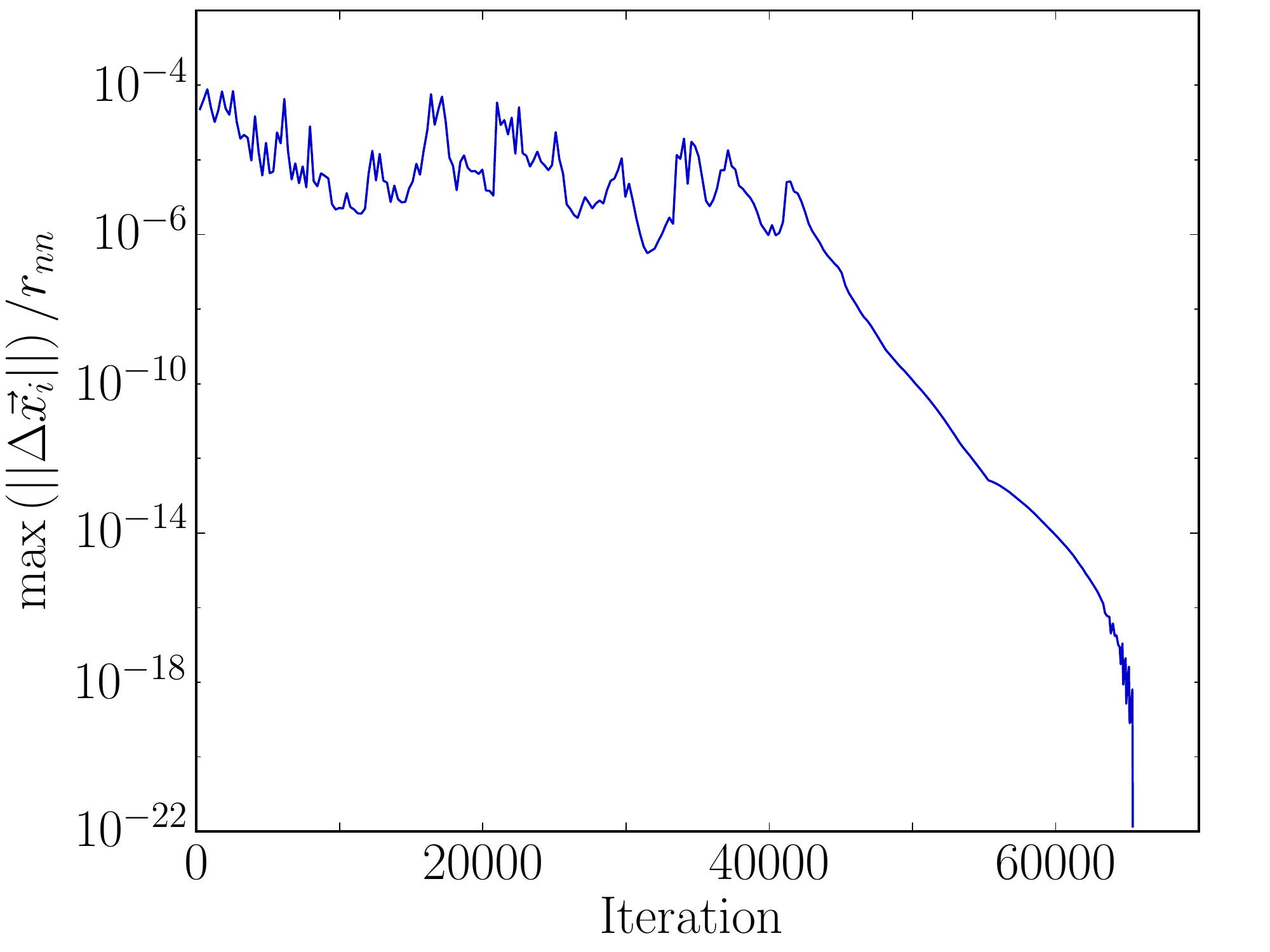}
\caption{Maximum particle displacement in a 3D packing with $N=1024$ particles, initally prepared in monodisperse equilibrium at $\varphi=0.7$ as we apply our iterative rescaling protocol. Each iteration consists of a Voronoi rescaling followed by a partial energy minimization. This system shows typical qualitative behavior for systems that converge to a fixed point.}
\label{fig:convergence}
\end{figure}

We track the dynamics of iteration by measuring the maximum square displacement, which provides an indication of this convergence. Sufficiently far from the jamming point, we find the behavior shown in figure \ref{fig:convergence}: systems quickly enter a regime of random motion within a consistent range of maximum square displacements. Eventually, the system (usually) falls into an approximately exponential descent until the displacement becomes impossible to resolve numerically. This reduction in displacement is connected to the existence of a fixed point, and the exponential behavior is related to the local analyticity of the energy landscape about that fixed point. Some systems instead remain in a state of random motion for a very long time, and limit cycles, while rare, have been observed. 

\begin{figure}[h]
\includegraphics[width=1\linewidth]{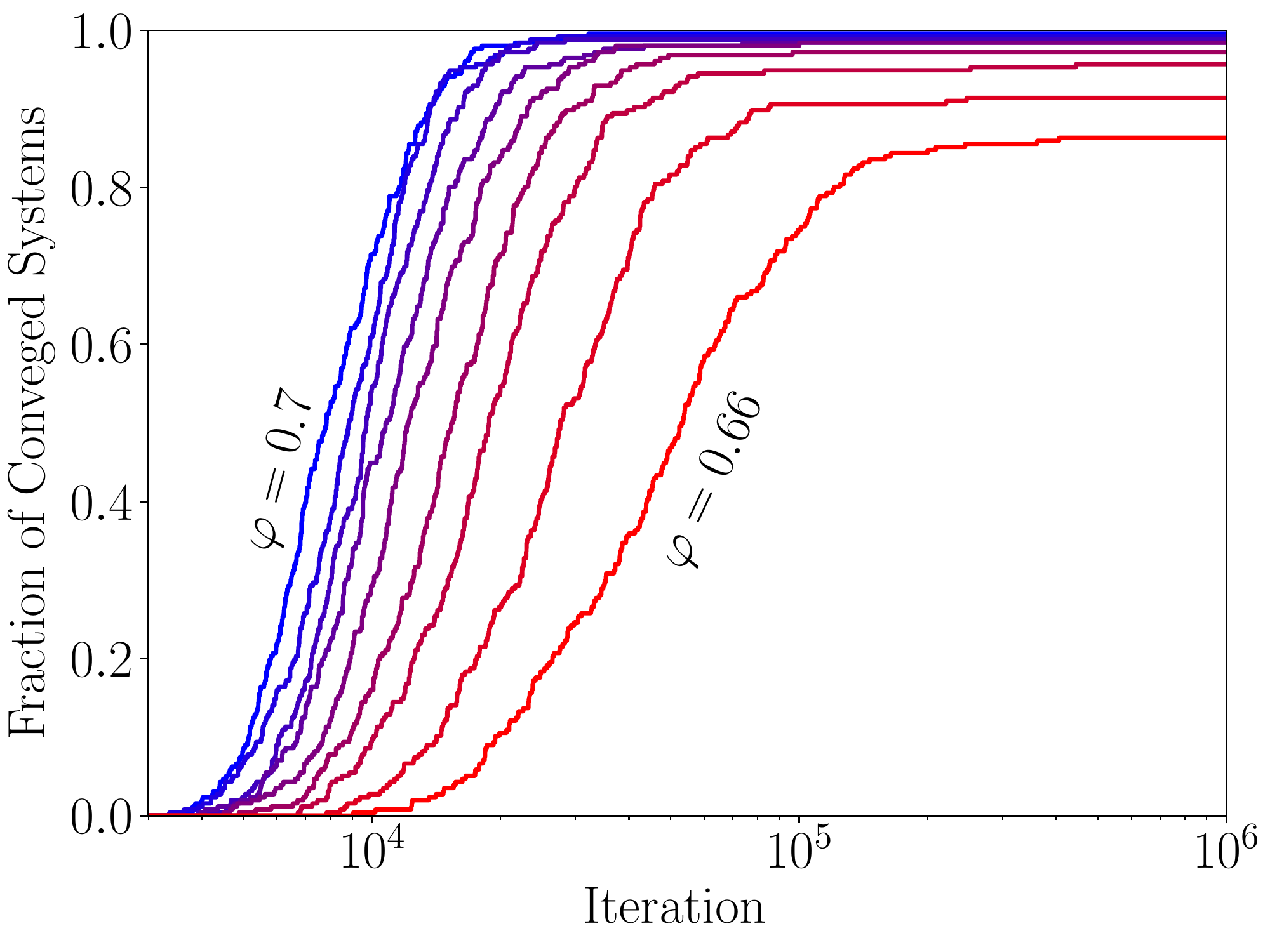}
\caption{Cumulative distributions of converged systems for a sampling of packings with $N=256$ particles at a range of packing fractions from $\varphi=0.7$ to $\varphi=0.66$, approaching the jamming density $\varphi_j \approx 0.64$. Packings nearer to jamming converge less frequently, and those that do converge do so more slowly. Larger systems show qualitatively similar behavior.}
\label{fig:convergenceCDF}
\end{figure}

Over an ensemble of many similar realizations, we observe that the probability of convergence approaches its maximum exponentially. This can be understood by assuming that falling into a fixed point is a Poisson process on the random walk through configuration space. The long-time probability of convergence depends on the parameters of the realizations, and most sensitively on the packing fraction, which we illustrate in figure \ref{fig:convergenceCDF}. While the characteristic time of convergence increases mildly as the jamming transition $\varphi_j$ is approached, the total probability of convergence decreases, carving out a regime around the critical point in which locating fixed points is at best computationally infeasible and at worst impossible. In order to ensure relatively quick convergence to fixed points, we primarily focus on systems relatively far from jamming.

\subsection{Structure Factor Scaling}

\begin{figure}[h]
\includegraphics[width=1\linewidth]{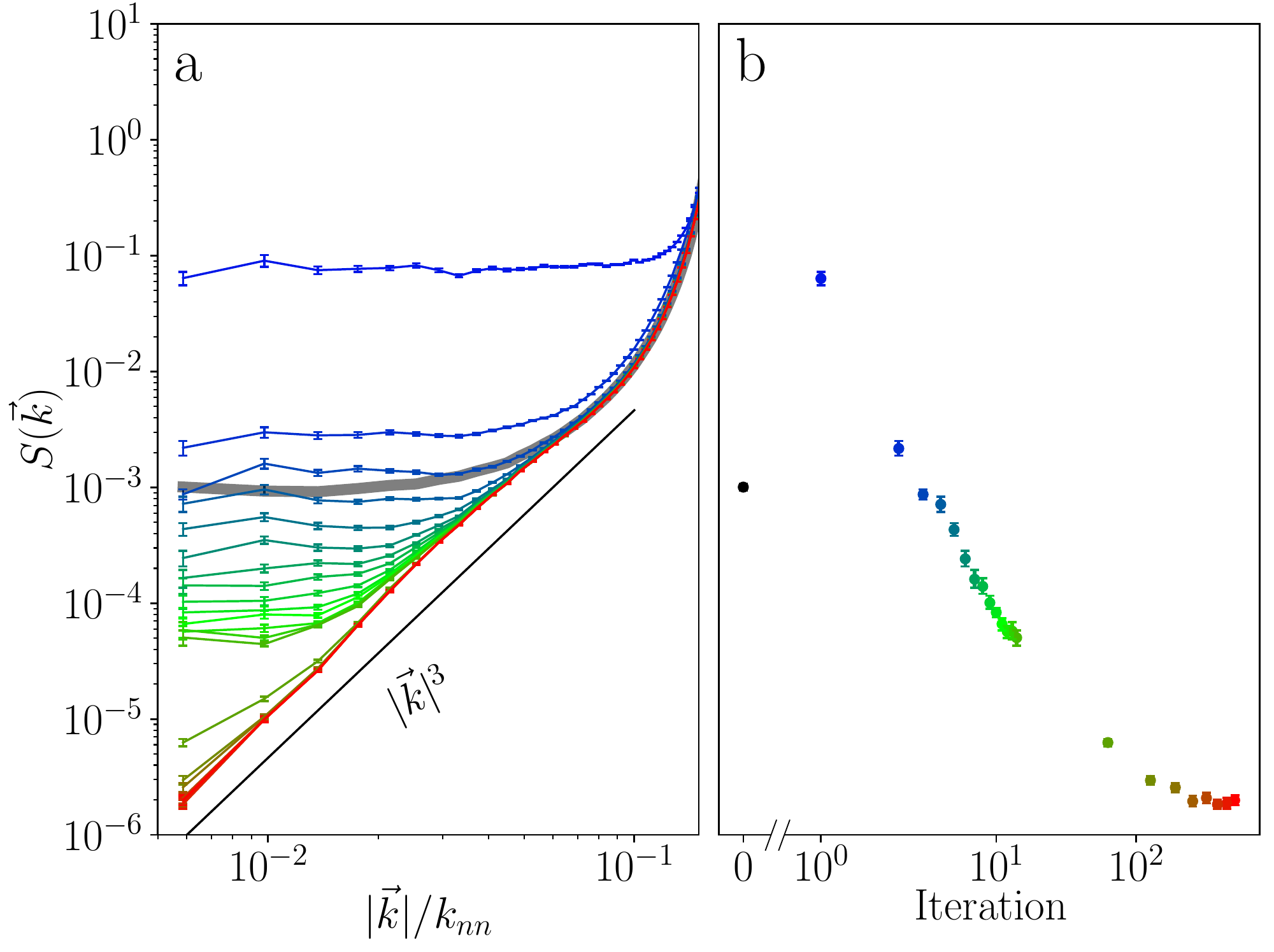}
\caption{a) Averaged structure factor $S(\vec{k})$ shown against wavevector magnitude $|\vec{k}|$ scaled by the nearest neighbor Poisson wavevector $k_{nn}$ for 13 packings of $N=65536$ particles in $d=3$. Packings are prepared initially in monodisperse equilibrium at $\varphi=0.8$. The structure factor for the initial configurations is shown as a grey solid line. As we iterate our rescaling protocol, the average structure factor is shown to evolve, (colors ranging from blue to green to red). b) $S(\vec{k}_\textrm{min})$ shown against the number of iterations of our rescaling protocol.}
\label{fig:structureFactor}
\end{figure}

The static structure factor $S(\vec{k})$ measures the strength of density correlations at a given wavevector $\vec{k}$~\cite{torquato_local_2003}. Any point pattern will have large correlations at short distances (large $\vec{k}$).  Uniform point patterns will have finite correlations even at very large distances ($\vec(k)$ approaching zero) whereas a hyperuniform point pattern will have vanishing correlations~\cite{torquato_local_2003}.

In Figure \ref{fig:structureFactor}, we show that iteration of our constructive algorithm suppresses long-range density fluctuations, as measured by the average value of $S(\vec{k})$ for a set of packings. Of particular interest is $\lim_{|\vec k|\to 0 }S(\vec k)$ which must be zero when we reach a hyperuniform fixed point. In the initial mechanically stable packing we observe that $S(\vec k)$ reaches a constant value as $\vec{k}$ approaches zero, indicating that it is not hyperuniform.  We consider the value of $S(\vec{k})$ at $\vec{k}_{\textrm{min}}$. After a single iteration this value of $S(\vec{k}_{\textrm{min}})$ increases, because individual iterations only perform partial minimizations, which initially introduces density fluctuations. As iteration continues, however, $S(\vec{k}_{\textrm{min}})$ decreases, indicating that the system is becoming hyperuniform.  Near the fixed point, the functional form of the structure factor resembles a power law $S(\vec k)\approx a|\vec k|^3$.

%The behavior of $S(\vec k)$ reveals that the dynamics of energy minimization \textit{do} influence structure at diverging length scales. Even if the low-$k$ uptick in the structure factor is removed by inducing hyperuniformity with a single rescaling, subsequent minimization will reliably reintroduce the behavior for all but a special class of systems. This suggests the low-$k$ behavior of stable soft sphere configurations observed here and elsewhere is not merely a result of initial configuration but a more generic feature of the model.

\subsection{Mechanical Properties}

\begin{figure}[h]
\includegraphics[width=1\linewidth]{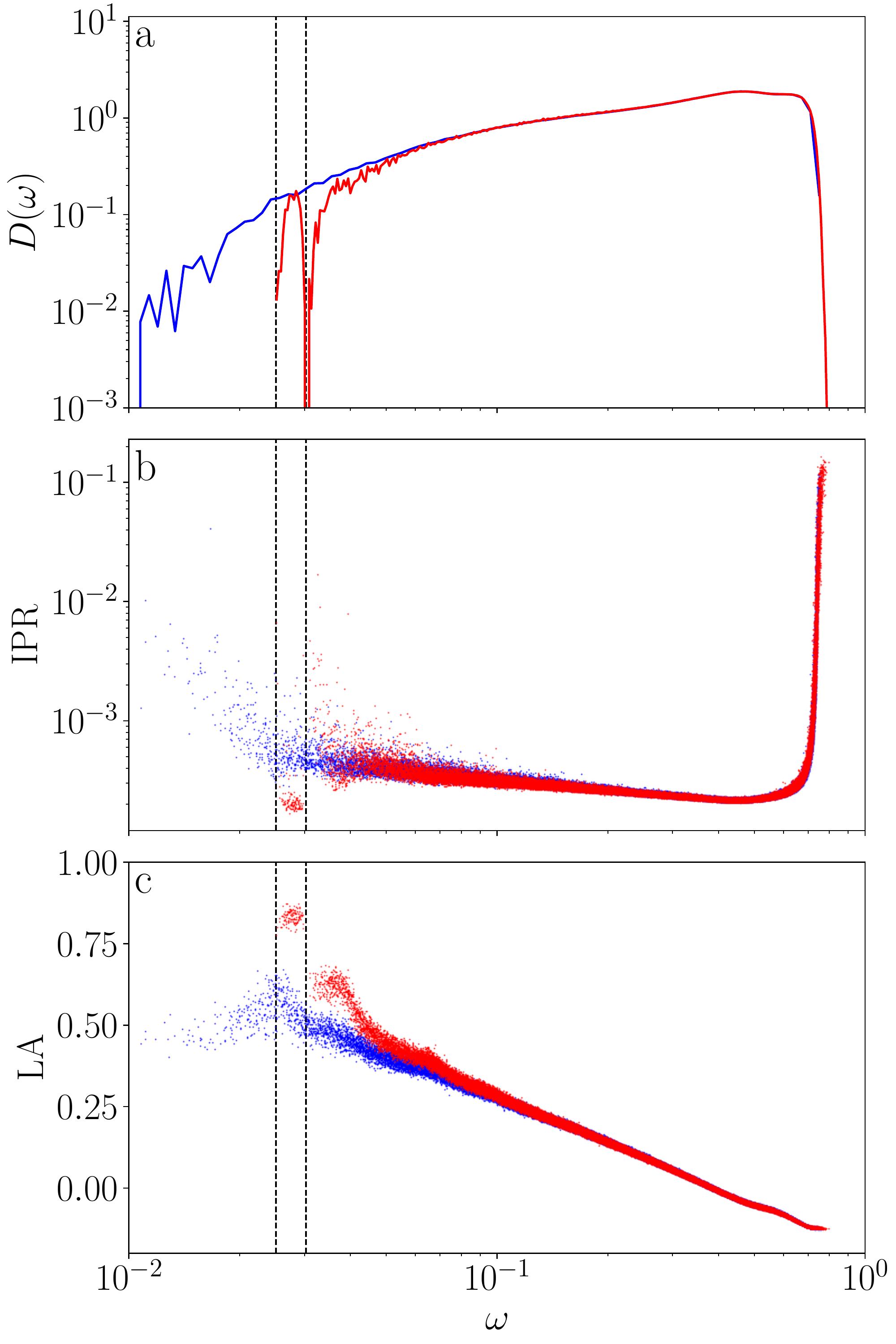}
\caption{Mechanical properties shown against frequency of vibrational modes $\omega$ for initial packings (blue) and hyperuniform final configurations (red). Sloshing modes, outlined in dashed grey lines, are shown to be easily distinguishable and characteristically different from the modes of a typical packing. Data is averaged over 22 packings with $N=8192$ in $d=3$ at $\varphi=0.7$. 
a) Density of States $D(\omega)$ shows the band gap around zero frequency and the band of sloshing modes in hyperuniform packings.
b) Inverse Participation Ratio (IPR), shows that sloshing modes are highly collective.
c) Local Affinity (LA) shows that sloshing modes are also highly correlated.}
\label{fig:mechanicalProperties}
\end{figure}

\begin{figure}[h]
\includegraphics[width=\linewidth]{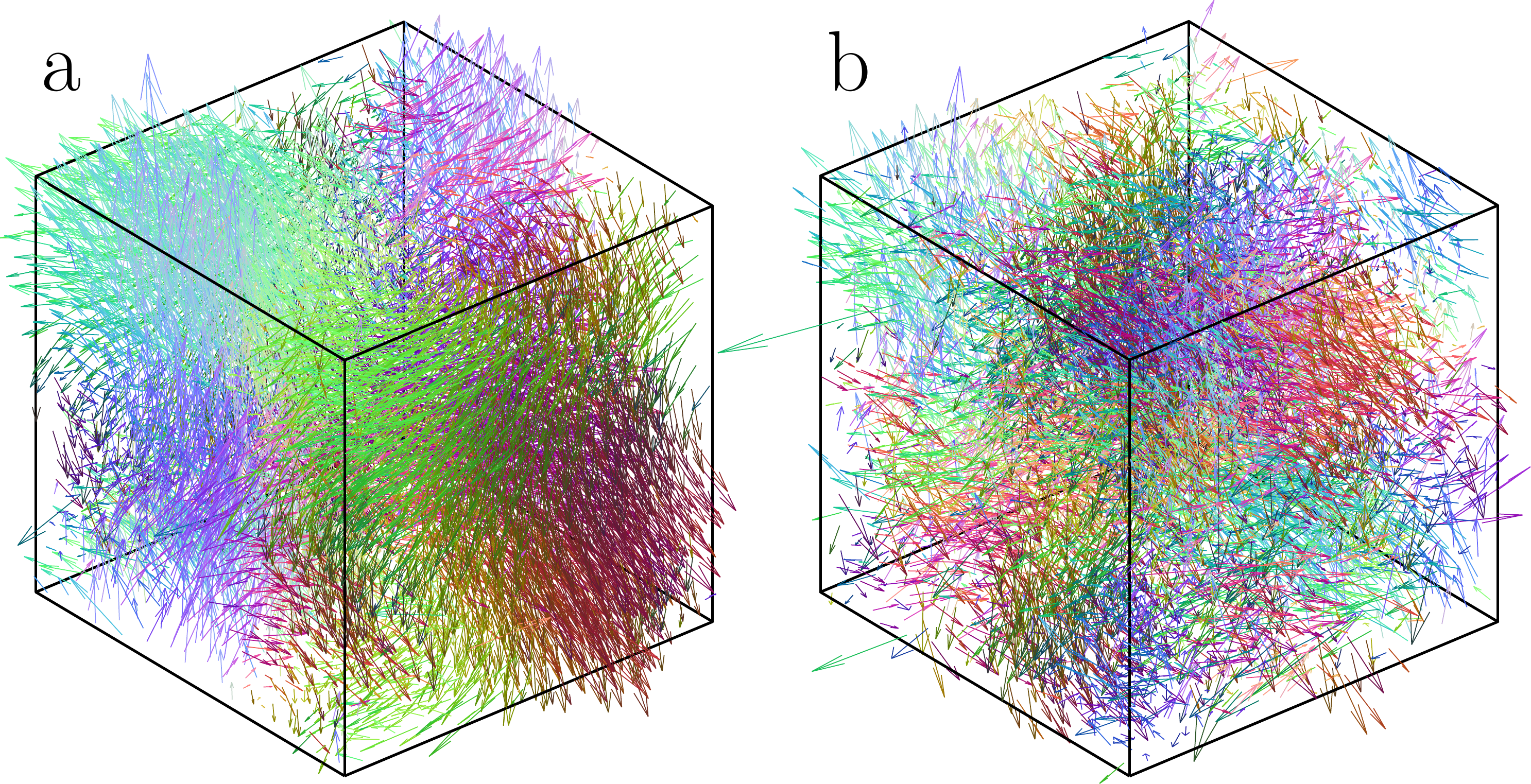}
\caption{Displacement field a typical sloshing mode in a hyperuniform packing (a) and a similar frequency mode in the initial packing (b). The length and color of each arrow represent the magnitude and direction of each particle's displacement in this mode. The normal mode of the hyperuniform packing is spatially coherent and reminiscent of a gyroidal structure, whereas that of the initial packing is incoherent.}
\label{fig:modeQuiver}
\end{figure}

%The connection between hyperuniform geometry and the mechanical properties of granular materials remains largely unexplored. 
Packings generated by Voronoi iteration represent a concrete set of examples with which to explore the mechanical properties of hyperuniform granular materials. The vibrational density of states $D(\omega)$ for sphere packings above jamming has been well characterized in previous work \cite{charbonneau_universal_2016, zhang_experimental_2017, arceri_vibrational_2020, mizuno_continuum_2017}. In figure \ref{fig:mechanicalProperties}(a), we show the density of states for both the initial packing and the hyperuniform final configuration. At the densities explored in this work we observe, as expected, a low frequency plateau associated with the Boson peak and then an $\omega^2$ falloff at the lowest frequencies in our initial packings. The final hyperuniform configurations, however, have a remarkably different density of states: the lowest frequency scaling regime is entirely absent, replaced by an abrupt low frequency cutoff at a finite frequency forming a phononic band gap about zero frequency. Above this gap, there is a discrete band of modes which we refer to as ``sloshing modes''.  The high frequency behavior, however, remains largely unchanged.

Figure \ref{fig:mechanicalProperties}(b) shows the inverse participation ratio (IPR) as a function of frequency, defined for a given eigenvector as
\begin{equation}
\text{IPR}=\frac{\sum_{i}\|\vec v_i\|^4}{\left(\sum_i\|\vec v_i\|^2\right)^2},
\end{equation}
where $\vec{v}_i$ is the mode displacement of the $i^\textrm{th}$ particle. This quantity measures the localization of each mode. The modes that constitute the sloshing band are easily distinguished from the initial configurations as they are highly collective, with typical IPR lower than at any other point in the spectrum.  The IPR for frequencies above the sloshing band is quantitatively similar to that observed in the initial configurations.

In addition to being collective, these modes are highly spatially correlated, consisting of nearly affine ``sloshing'' motions over large portions of the system, visualized in Figure \ref{fig:modeQuiver}. To quantify this, we define the \textit{local affinity} (LA) of a single mode $\vec v_i$ as
\begin{equation}
\text{LA}=\frac{1}{N_{\text{neigh}}}\sum_{\langle i,j\rangle}\frac{\vec v_i\cdot\vec v_j}{\|\vec v_i\|\ \|\vec v_j\|},
\end{equation}
where $i$ indicates the particle index and $\langle i,j\rangle$ denotes the $N_\text{neigh}$ neighbors in the radical Delaunay triangulation~\cite{aurenhammer_voronoi_2013}. For each mode this quantity measures the extent to which adjacent particles oscillate in phase and in the same direction, having a value of 1 for a spatially affine displacement of the entire system. Figure \ref{fig:mechanicalProperties}(c) shows that when we measure the local affinity, we again find a clear distinction between the sloshing modes in the hyperuniform systems and the much less correlated low frequency modes of the initial configurations. Modes within the sloshing band are consistently more affine then the rest.

\section{Conclusions}

We have demonstrated that hyperuniform, overjammed, polydisperse granular packings not only exist, but can be efficiently generated at large system size through a process of iterative optimization. This result allows us to probe a concrete set of systems to better understand the implications of hyperuniformity.

Despite being chosen on geometrical grounds, the hyperuniform packings we construct exhibit unusual mechanical properties. We have investigated these by exploring their linear vibrational modes. Here we find a low energy band gap structure not seen in typical preparations, which includes highly collective excitations and regions of reduced density of states, indicative of the formation of phononic band gaps. Further modification of the band structure of granular media may be possible through similar geometric optimization.

Additionally, the lack of normal modes at low frequency suggests that these packings are highly stable against mechanical perturbation. Accordingly we believe that they lie in particularly deep wells in the soft sphere energy landscape. This suggests that the procedure of smoothing out density fluctuations through Voronoi iteration may be somewhat analogous to aging in thermal glasses. Further elaboration of this correspondence could shed light on the geometric properties of aged glasses, and the role of large-scale geometric heterogeneity in glassy dynamics.

\section{Acknowledgements}
We thank Ludovic Berthier, Raghuveer Parthasarathy, and Taiki Yanagishima for valuable discussion and feedback. This work is supported by the Simons Collaboration on Cracking the Glass Problem via award No. 454939.

\bibliography{VoronoiFixedPointBib}

\end{document}